\documentstyle[epsfig,aps,prc]{revtex}
\begin{document}

%\begin{titlepage}

\title{Correlation effects on the static structure factor
of a Bose gas}

\author{Ch.C. Moustakidis}
\address{ Department of Theoretical Physics,
Aristotle University of Thessaloniki,
GR-54124 Thessaloniki, Greece}

\maketitle
\begin{abstract}
A theoretical treatment of the  static structure factor $S(k)$ of
a Bose gas is attempted. The low order expansion theory is
implemented  for the construction of the two body density
distribution, while various trial functions for the radial distribution
function $g(r)$ are used. $g(r)$ introduces the atomic correlations and
describes the departure from the noninteracting gas.
The Bose gas is studied as inhomogeneous one, trapped in harmonic
oscillator well, as well as homogeneous. A suitable parametrization
of the various trial functions $g(r)$ exists which leads to satisfactory
reproduction of the experimental  values of $S(k)$, both in
inhomogeneous case as well as in homogeneous one.
The phonon range behavior of the calculated $S(k)$ is also
addressed and discussed both in finite and infinite Bose gas.

\vspace{5mm}
\vspace{3mm}
\end{abstract}
%\end{titlepage}
%\vspace*{2cm}

%%%%%%%%%%%%%%%%%%%%%%%%%%%%%%%%%%%%%%%%%%%%%%%%%%%%%%%%%%%%%%%%%%%%%%%%%%%%%%%%%%%%%%%%%%%%%%%%%%%%%%%%%%%%%%%%%%
%%%%%%%%%%%%%%%%%%%%%%%%%%%%%%%%%%%%%%%%%%%%%%%%%%%%%%%%%%%%%%%%%%%%%%%%%%%%%%%%%%%%%%%%%%%%%%%%%%%%%%%%%%%%%%%%%%
%\vspace*{1cm}
\section{Introduction}

Spectroscopic studies have been used to assemble a complete
understanding of the structure of atoms and simple molecules
\cite{Stamper-Kurn}. The static structure factor $S(k)$ is a
fundamental quantity, which is connected with the atomic
structure, and it is the Fourier transform of the radial distribution
function $g(r)$. $S(k)$ gives the magnitude of the density
fluctuation in the system (atomic, molecular, electronic or
nuclear system) at wavelength $2\pi/k$, where $k$ is the momentum
transfer. In  recent papers, the Bragg spectroscopic method was
used  to measure $S(k)$ either in the phonon regime
\cite{Stamper-Kurn} or/and in the single-particle regime
\cite{Steinhauer}.

More specifically, the character of excitations in a weakly
interacting Bose-Einstein condensed gas depends on the relation
between the wave vector of the excitation $k$ and the inverse
healing length $ \xi^{-1}=\sqrt{2}mc_s/\hbar=\sqrt{8\pi a \rho}$,
which is the wave vector related to the speed of Bogoliubov sound
$c_s=\sqrt{\mu/m}$, where $\mu=4\pi \hbar^2 a\rho/m$ is the
chemical potential, $a$ is the scattering length, $\rho$ is the
condensate density, and $m$ is the atomic mass
\cite{Stamper-Kurn,Fetter98}. There are two different regimes of
excitation described by the Bogoliubov theory for the
zero-temperature weakly interacting Bose-Einstein condensate: the
phonon and the free-particle regimes. For small wave vectors ($k
\ll \xi^{-1}$), the gas responds collectively and density
perturbations propagate as phonons at the speed of Bogoliubov
sound. For large wave vectors ($k \gg \xi^{-1}$) the excitations
are particle-like with a quadratic dispersion relation
\cite{Stamper-Kurn,Fetter98}.

In the present letter we focus on the theoretical calculation of
$S(k)$ of a Bose gas. All the calculations are
performed in the zero-temperature limit ($T=0$). Various types of
trial functions $g(r)$ are used and the entailed $S(k)$ are
compared with the experimental data. The correlation parameters of
$g(r)$ are determined by fit to the experimental data. The theoretical
investigation is performed by applying two approaches. In the
first approach, the Bose gas is treated as inhomogeneous gas, which is
exactly the case of the  experiments
\cite{Steinhauer,Stamper-Kurn}. We consider that the  Bose gas is
trapped in a harmonic oscillator well with condensate radius $R$.
In the second approach, the trapped bose gas is treated, in a
rough approximation, as homogeneous gas in regard to its elementary
excitations. This approximate treatment is reasonable in the
sense that the coherence length $\xi$ satisfies the condition  $\xi
\ll R$ \cite{Graham}. Actually in the experiments of Refs.
\cite{Stamper-Kurn,Steinhauer} this condition is well satisfied.

As mentioned before, the key quantity of the above approaches,
is the radial distribution function $g(r)$, which describes the
relative probability for finding two particles at a distance $r$
apart and consequently it introduces the atomic correlations.
In general, the function
$g(r)$ can be calculated by the variational method
\cite{Mazzanti}. However, in dilute Bose gas, in the framework of
the lowest-order cluster expansion,  trial
expressions for  $g(r)$ can be used
\cite{Mazzanti,Jastrow,Steinhauer04} and
the parameters can be fitted to reproduce the experimental data of
$S(k)$. Up to now,  $S(k)$ in a trapped Bose gas, is  predicted
successfully (compared to the experimental data) by the local
density approximation \cite{Zambelli,Steinhauer}.

The motivation of the present work is the theoretical study of
$S(k)$, considering the Bose gas as a many-body system. It is well
known that the ideal Bose gas model (noninteracting model) fails to
reproduce the experimental values of $S(k)$. We focus our
treatment on the effect of interatomic interaction on the
properties of $S(k)$, both in small and large values of the
momentum transfer $k$. It is fundamental to investigate how  these
effects modify the picture of $S(k)$ of an ideal gas and to
predict values  consistent with the experimental data. The
possibility of linear behavior of $S(k)$ for small $k$, both in
inhomogeneous and homogeneous Bose gas has also been examined.
In section 2 we treat the inhomogeneous Bose gas while the case of
homogeneous Bose gas is investigated in section 3.

%\section{The model}
%%%%%%%%%%%%%%%%%%%%
%%%%%%%%%%%%%%%%%%%%
\section{Inhomogeneous Bose gas}

A dilute inhomogeneous Bose gas can be studied
using the low-order approximation (LOA)
\cite{Mazzanti,Jastrow,Steinhauer04}.
In the LOA the two-body density distribution (TBDD) has the form
\begin{equation}
\rho({\bf r}_1,{\bf r}_2)= C \rho({\bf r}_1)\rho({\bf r}_2)
f^2(r_{12})=C \rho({\bf r}_1)\rho({\bf r}_2)g(r_{12}) ,
\label{TBDD-1}
\end{equation}
where $f(r_{12})$ ($r_{12}=r=|{\bf r}_1-{\bf r}_2|$) is the
Jastrow correlation function \cite{Jastrow}, $ g(r_{12})$ is the
radial distribution function, $C$ is the normalization factor which
ensures that $\int \rho({\bf r}_1,{\bf r}_2) d{\bf r}_1
d{\bf r}_2=N(N-1) $ (N is the number of the atoms of the Bose
condensate) and
$\rho({\bf r})$ is the density distribution (DD) of the system.
The TBDD is proportional to the probability of simultaneously
finding an atom at ${\bf r}_1$ and another at ${\bf r}_2$. In
the uncorrelated case (noninteracting gas) the TBDD takes the
simple form
\begin{equation}
\rho({\bf r}_1,{\bf r}_2)=\frac{N-1}{N} \rho({\bf r}_1)\rho({\bf r}_2).
\label{TBDD-2}
\end{equation}

In the present work we consider that the atoms are confined in a harmonic
oscillator trap where the ground state single-particle  wave
function $\psi_o({\bf r})$  has the form
\[
\psi_{o}(r)=\frac{N^{1/2}}{\pi^{3/4} b^{3/2}}
\exp [-r^2/(2 b^2)], \qquad b=[\hbar/(m\omega)]^{1/2}. \]
The normalization of the DD is
$\int \mid \rho({\bf r}) \mid^2  d {\bf r}=
\int \mid \psi_0({\bf r}) \mid^2 d {\bf r} =N$.

The static structure factor $S(k)$,
in a finite system, is defined as \cite{Zambelli}
\begin{equation}
S({\bf k})=1+
\frac{1}{N} \int e^{i{\bf k}({\bf r}_1-{\bf r}_2)}
\left[ \rho({\bf r}_1,{\bf r}_2)
- \rho({\bf r}_1) \rho({\bf r}_2 ) \right]
d  {\bf r}_1   d  {\bf r}_2
\label{str-fin}
\end{equation}
or using Eq. (\ref{TBDD-1})
\begin{equation}
S({\bf k})=1+
\frac{1}{N} \int e^{i{\bf k}({\bf r}_1-{\bf r}_2)}
\rho({\bf r}_1) \rho({\bf r}_2 )
[C g(r_{12})-1]
d  {\bf r}_1   d  {\bf r}_2
\label{str-fin2}
\end{equation}

The integration in Eq. (\ref{str-fin2}) can be performed if the
function $g(r)$ is know. $g(r)$ must obey  the rules $g(r=0)=0$
and $\lim_{r \rightarrow \infty} g(r) \rightarrow 1$. The first
rule introduces the repulsive correlations between the atoms and
the second the absence of such correlations in long distances. In
general the form of $S(k)$ is affected appreciably from the form
of $g(r)$. More specifically, the long range behavior of $g(r)$
affects $S(k)$ for small values of $k$ while the short range
behaviour of it affects $S(k)$ for large values of $k$ as a direct
consequence of the Fourier transform theory.

In the present work, we choose two trial forms for $g(r)$. The
first one is a gaussian type  which has been extensively and
successfully used for the study of similar problems in atomic
physics (Bose gas, liquid helium) as well in nuclear physics. The
relevant $g(r)$ and the entailed $S(k)$ are
\begin{eqnarray}
{\rm Case \ 1} \qquad & & \nonumber \\
g(r)&=& 1- \exp[-\beta r^2]\nonumber \\
S(k)&=&1+N(C_1-1)\exp[-k_b^2/2]
-\frac{N C_1}{(1+2y^2)^{3/2}}\exp[-k_b^2/2(1+2y)],
\label{fin-cs-1}
\end{eqnarray}
where $k_b=kb$, $y=\beta b^2 $, $\beta$ is the correlation parameter
and $C_1$ is the normalization factor.

The second trial function $g(r)$ and the relevant $S(k)$ are of the form
\begin{eqnarray}
{\rm Case \ 2} \qquad & & \nonumber \\
g(r)&=& 1- \frac{\sin^4ar}{(ar)^4}\nonumber \\
S(k)&=&1+N(C_2-1)\exp[-k_b^2/2] \nonumber\\
&-& \frac{N C_2}{2^{15/2} ab \pi k_b}
\sum_{i=1}^{5} \alpha_i\left[\beta_i\exp[-\beta_i^2/4]+
\sqrt{\pi}(1+\frac{\beta_i^2}{2}) {\rm erf}(\frac{\beta_i}{2}) \right]
\label{fin-cs-2}
\end{eqnarray}
where $a$ is the correlation parameter,
$\alpha_i$ are known  coefficients, $\beta_i=\beta_i(kb,a,b)$,
${\rm erf}(z)=2/\sqrt{\pi} \int_{0}^{z} e^{-t^2} dt$ and
$C_2$ is the normalization factor.

Trial functions $g(r)$ of hard-sphere form ($g(r)=1-a/r$) and soft
repulsive form ($g(r)=1-\sin ar/ar$) are also used. Both of them,
lead either to the ideal Bose gas results (negligible
correlations) or to abnormal fluctuating negative and positive
values for $S(k)$ (strong correlations) and consequently in
complete disagreement with the experimental data
\cite{Moustakidis04}. Therefore, it should be emphasized that the
choice of a trial function $g(r)$  should be made  under certain
restrictions in order to get reasonable values for  $S(k)$.

The behavior of S(k), in case 1, for different values of the
correlation parameter $\beta$ is shown in Fig. 1(a). It is obvious
that the effect of correlations, induced by the function
$g(r)$, becomes large when the parameter $\beta$ becomes small and
vice versa. The case where $\beta\rightarrow \infty$, corresponds
to the uncorrelated case (HO). For the values of $k$ employed in
the experiment of Ref. \cite{Steinhauer} (hereafter EXP-1) the
prediction of the HO model is always close to 1 for $S(k)$. When
the correlation parameter $\beta$  decreases considerably
(strong correlations) the theoretical prediction of $S(k)$ is in
good agreement with the experimental data. The value $\beta=5.3 \
\mu m^{-2}$ gives the best least squares fit in that case. In
general the gaussian form of $g(r)$, in spite of its simplicity,
reproduces fairly well the experimental data of EXP-1, both in
low and high values of the momentum $k$. It reproduces also the
experimental data of Ref. \cite{Stamper-Kurn} (hereafter EXP-2) as
can been seen from Fig. 1(c). Within our theoretical model, the
gaussian type of $g(r)$ is flexible enough to obtain values for
$S(k)$ in agreement with the experimental data.

Fig. 1(b) displays the results in case 2, which are
compared with those of the data of
EXP-1. The model reproduces well the
experimental data in the range $1.5-3 \ \mu m^{-1}$ (with best least squares
fit value $a=1.34 \  \mu m^{-1}$), but fails in
the range $k>3 \  \mu m^{-1}$. The main drawback of this model is the
predicted negative values of  $S(k)$ in the range
close to $k=0$ when the
correlation parameter $a$ decreases considerably (strong
correlation case).

The correlation function $g(r)$ corresponding to
cases 1 and 2 for the correlation parameters $\beta=5.3 \ \mu m^{-2}$ and
$a=1.34 \ \mu m^{-1} $ respectively is sketched in Fig. 2(a).
Those values of the parameters $\beta$ and $a$ give the best $x^2$
value in the fit of the theoretical expressions of $S(k)$ to the
data of EXP-1.
The most striking feature in case 2 is the
existence of strong correlations, introduced by  $g(r)$,
in order to reproduce the experimental data of $S(k)$.
It is worthwhile to point out that $g(r)$, in case 2,  exhibits
fluctuations in the range $r>2 \  \mu m$ but this is not visible in Fig.
2(a).

The possibility of a linear dependence of $S(k)$ on $k$ for small values
of $k$, as predicted from other works \cite{Zambelli},
is prohibitive, on the basis of Eq. (\ref{str-fin})
at least in the case where the trap is an harmonic oscillator one.
That can be seen considering  the ground state wave function
to be the harmonic oscillator one and transforming ${\bf r}_1$
and ${\bf r}_2$ in Eq. (\ref{str-fin}) into the coordinates
of the relative motion (${\bf r}={\bf r}_1-{\bf r}_2$) and the center of
mass motion (${\bf R}=({\bf r}_1+{\bf r}_2)/2 $).
After some algebra $S({\bf k})$ takes the form
\begin{equation}
S({\bf k}) \sim \int e^{i{\bf k}{\bf r}} e^{-r^2}
[Cg(r)-1] d  {\bf r}
\label{sk-rel}
\end{equation}

For finite systems, as is a trapped Bose gas,
we can expand the
exponential $e^{i{\bf k}{\bf r}}$, since ${\bf r}$ is bounded.
So that:
\begin{equation}
e^{i{\bf k}{\bf r}}=1+i{\bf k}{\bf r}
+\frac{(i{\bf k}{\bf r})^2}{2 !}+
\frac{(i{\bf k}{\bf r})^3}{3 !}+\cdots
\label{expand}
\end{equation}

Substituting  Eq. (\ref{expand}) into Eq. (\ref{sk-rel}) and
considering that the terms with odd powers of $k$ do not
contribute on the integral, $S(k)$ takes the form
\begin{equation}
S(k) \sim  a_1k^2+a_2k^4+\cdots
\label{sk-rel-2}
\end{equation}

Thus, for small values of $k$, $S(k)$ depends linearly on $k^2$.
The gaussian factor $e^{-r^2}$, originating from the harmonic
oscillator wave function of the trapped Bose gas, ensures the
convergence of the integrals $a_i$ corresponding  to the even powers
of the expansion.

%%%%%%%%%%%%%%%%%%%%%%%%%%%%%%%%%%%%%%%%%%%%%%%%%%%%%%%
%%%%%%%%%%%%%%%%%%%%%%%%%%%%%%%%%%%%%%%%%%%%%%%%%%%%%

\section{Homogeneous Bose gas}

The condensate of an inhomogeneous (finite) Bose gas, can be
treated as homogeneous in regard to its elementary excitations
considering that the coherence length $\xi$ satisfies $\xi \ll R$
\cite{Graham}. Actually, this is a rough approximation, but can
effectively describe fairly well  the excitation properties of a
trapped Bose gas.

In infinite systems $\rho({\bf r}_1)$ is
constant ($\rho({\bf r}_1)=\rho$) and thus the TBDD is given by $
\rho({\bf r}_1,{\bf r}_2)=\rho^2 g(r_{12}) $. Thus, the  structure
factor of homogeneous gas (or quantum fluid in general) is given
by the relation \cite{Feenberg}
\begin{equation}
S({\bf k})=1+\rho \int e^{i{\bf k}{\bf r}} [g(r)-1] d  {\bf r},
\label{str-ifin-1}
\end{equation}
or after performing the angular integration
\begin{equation}
S(k)=1+4 \pi \rho \int_{0}^{\infty} \frac{\sin kr}{kr} r^2 [g(r)-1]  dr.
\label{str-ifin-2}
\end{equation}
The necessary conditions which must be obeyed by $g(r)$ and $S(k)$, when a
given $g(r)$ is not generated directly and explicitly by a wave
function, include \cite{Feenberg}
\begin{eqnarray}
&& g(r)\geq 0, \qquad S(k)\geq 0, \qquad
\lim_{k \rightarrow 0} S(k)=\frac{\hbar k}{2mc},
\nonumber\\
&& S(0)=0 \ \Rightarrow \ \
4 \pi \rho \int_{0}^{\infty} r^2 [g(r)-1]  d r=-1
\label{conditions}
\end{eqnarray}
The third condition of (\ref{conditions}) imposes the condition for
the phonon like excitation on the Bose gas and is satisfied if and only if
\cite{Feenberg}
\begin{equation}
\lim_{r \rightarrow \infty} r^4[\overline{g(r)}-1]=
\frac{-\hbar}{2\pi^2 m \rho c_s}
\label{beh1/r4}
\end{equation}
where the bar denotes an average over a range $\delta r$ somewhat larger
than $\rho^{-1/3}$ \cite{Feenberg}.

We have used three different trial expressions for $g(r)$ to study
$S(k)$ in a uniform Bose gas. The first two were used already in the study of
the inhomogeneous Bose gas. However, the relevant structure factor,
as expected, has different forms  compared to inhomogenous case.
The used forms of $g(r)$ and the entailed $S(k)$ are
\begin{eqnarray}
{\rm Case \ 3} & & \qquad \nonumber \\
g(r)&=&1-\exp[-\beta r^2]
\nonumber\\
S(k)&=&1-\exp[-\frac{k^2}{4 \beta}]
\label{case-2}
\end{eqnarray}
%%%%%%%%%%%%%%%%%%%%%%%%%%%%%%%%%%%%%
\begin{eqnarray}
{\rm Case \ 4} \qquad & & \nonumber \\
g(r)&=&1-\frac{\sin^4 ar}{(ar)^4}
\nonumber\\
S(k)&=&\left\{ \begin{array}{cc}
\frac{3k}{8a}                 , &  \mbox{$k<2a$} \\
2-\frac{2a}{k}-\frac{k}{8a}   , &  \mbox{$2a<k<4a$}\\
1                             , &  \mbox{$4a<k$}
                              \end{array}
                       \right.
\label{case-5}
\end{eqnarray}
%%%%%%%%%%%%%%%%%%%%%%%%%%%%%%%%%%%%
Case 5 is a combination, in a way,
of cases 3 and 4 and takes the form
\begin{eqnarray}
{\rm Case \ 5} \qquad & & \nonumber \\
g(r)&=&1-\frac{\sin ar}{ar} \exp[-c r]
\nonumber\\
S(k)&=&\frac{k^2(k^2+2c^2-2a^2)}{(a^2+c^2)^2+k^2(k^2+2c^2-2a^2)}
\label{case-3}
\end{eqnarray}
%%%%%%%%%%%%%%%%%%%%%%%%%%%%%%%%%%%%%

It is worth  noticing that in homogeneous cases, due to the fourth
condition of Eq. (\ref{conditions}), the correlation parameters of
$g(r)$, in all the cases, depend directly on the density $\rho$.
Thus, the fourth condition of Eq. (\ref{conditions}), leads to the
relations $\rho=\beta^{3/2}/\pi^{3/2}$ (case 3), $\rho=a^3/\pi^2$
(case 4) and $\rho=(a^2+c^2)^2/8c\pi$ (case 5).

In the present work we compare also our results for  $S(k)$ with those
of the hard-sphere interaction
in uniform dilute gas, which was predicted long ago by
Lee et al. \cite{Lee57} and has the form
\begin{equation}
S(k)=\frac{k}{\sqrt{k^2+16 \pi \alpha \rho}}
\label{case -lee}
\end{equation}
where $\alpha$ is the hard-sphere diameter 
(see also Refs. \cite{Isihara57,Shanenko97}).

The values of $S(k)$ in case 3 (gaussian $g(r)$) for various
values of the constant density $\rho$ (or of the correlation parameter
$\beta$) are displayed in Fig.
3(a). As in the corresponding case of the inhomogeneous Bose gas
that type of $g(r)$ reproduces the data of both experiments, see
also Fig. 1(c), when the density of the gas is approximately 
$\approx 1 \  \mu m^{-3}=10^{12} \ cm^{-3}$. We note that in the
uncorrelated case (noninteracting model), the static structure
factor of homogeneous Bose gas has the constant value $S(k)=1$.
When the density of the Bose gas increases, the calculated values
of  $S(k)$ deviates from the experimental data, exhibiting
considerably lower values at the same range of the momentum
transfer $k$. Moreover, $S(k)$ behaves quadratically in the phonon
regime ($\lim_{k \rightarrow 0}S(k) \sim k^2$), in contradiction
with the condition (\ref{conditions}), as a result of the
improper long range behavior of $g(r)$ (see Eq. (\ref{beh1/r4})).

The behavior of  $S(k)$, in case 4, is illustrated  in Fig. 3(b).
A striking future in this case is the linear behavior of  $S(k)$
($\lim_{k \rightarrow 0}S(k) \sim k $), in the phonon regime, as a
consequence of the proper long range behavior of the function
$g(r)$. The best fit value of the correlation parameter
a corresponds, as in the previous case,
to density $\rho\sim 1 \  \mu m^{-3}=10^{12} \ cm^{-3}$.
Increasing the density, lower values of  $S(k)$ appear for the
same range of the momentum $k$.

$S(k)$ in case 5, is plotted in Fig. 3(c).
In this case $g(r)$, combines in a way  cases 3 and 4.
$S(k)$ behaves quadratically in phonon regime ($S(k \rightarrow 0) \sim
k^2$) but reproduces quite well the experimental data, especially
in the range $0-4 \  \mu m^{-1}$. $S(k)$, derived by Lee et
al. \cite{Lee57}, is also plotted in  Fig. 3(c). In that case, the phonon
behavior is ensured, while the prediction is in good
agreement with the experimental data.

The theoretical results of $S(k)$ in cases 1,3,4 and 5 using the
best fit values of the parameters are compared with the experimental
data of Exp-2 in Fig. 1(c). It is seen that in all  cases there is
a good agreement between the theoretical values and the experimental data.

The behavior of the correlation function $g(r)$, for  cases
3,4 and 5, is sketched in Fig. 2(b). The most striking feature is that
though the behavior of  $g(r)$ is almost the same in cases 3
and 4, the corresponding $S(k)$, display different behavior,
especially for the lower values of the momentum $k$. It is concluded
that  $S(k)$ is very sensitive to the form of  $g(r)$ and
specifically, the long range behavior of $g(r)$ affects considerably
the behavior of $S(k)$ in the phonon regime (small values of
$k$). This is a well known property of the Fourier transform.

It is worth remarking that a correlation function of the
hard-sphere form, $g(r)=1-a/r$, thought it is as realistic one, mainly for
low values of the interatomic distance $r$, 
leads to $S(k)=1-4\pi\rho a/k^2$ which diverges when $k\rightarrow 0$.
%is inapplicable
%because the integral of Eq. (\ref{str-ifin-1}) diverges
%\cite{Moustakidis04}.

In conclusion, we report a theoretical calculation of the static
structure factor $S(k)$ both for inhomogeneous and homogeneous
Bose gas, in the framework of the low order expansion theory,
by applying various trial forms for the radial distribution function
$g(r)$. We compared our results with recent experimental data
concerning trapped Bose gas. The correlation parameters of
$g(r)$ are  adjusted in order to reproduce the experimental data.
By applying suitable parametrization the experimental data are
reproduced quite well. The low $k$ behavior of the calculated
$S(k)$ is also addressed and discussed both in inhomogeneous
and homogeneous Bose gas.

%%%%%%%%%%%%%%%%%%%%%%%%%%%%%%%%%%%%%%%%%%%%%%%%%%%%%%%%%%%%%%
%%%%%%%%%%%%%%%%%%%%%%%%%%%%%%%%%%%%%%%%%%%%%%%%%%%%%%%%%%%%%%
\vspace{4mm}
%\begin{center}
%{\bf ACKNOWLEDGMENTS}
%\end{center}

The author would like to thank Prof. S.E. Massen for fruitful discussions
and comments on the manuscript, Prof. A.L. Fetter for valuable
comments and correspondence for the low limit (phonon) behavior of
the static structure factor,
Dr. C.P. Panos for useful comments on the manuscript,
Dr. J. Steinhauer for comments on the hard-sphere correlation
function,
N. Katz for literature information
about the experimental data for  $S(k)$ and Prof. A. Herrera-Aguilar
for fruitful comments concerning case 2.

%%%%%%%%%%%%%%%%%%%%%%%%%%%%%%%%%%%%%%%%%%%%%%%%%%
%%%%%%%%%%%%%%%%%%%%%%%%%%%%%%%%%%%%%%%%%%%%%%%%%%

%\end{document}

\newpage
%%%%%%%%%%%%%%%%%%%% Figure 1
\begin{figure}
\begin{center}
\begin{tabular}{c}
{\epsfig{figure=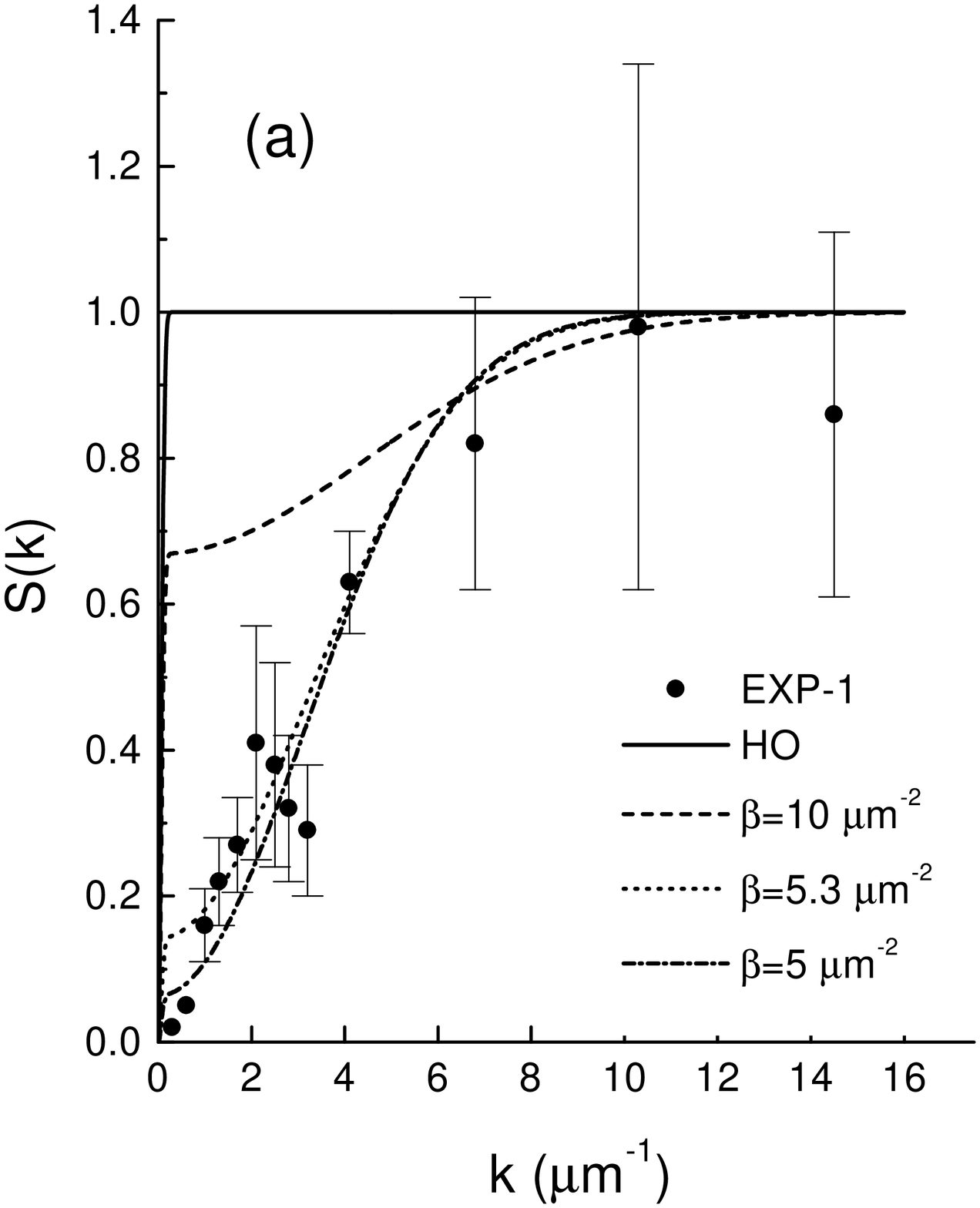,width=5.5cm} }
{\epsfig{figure=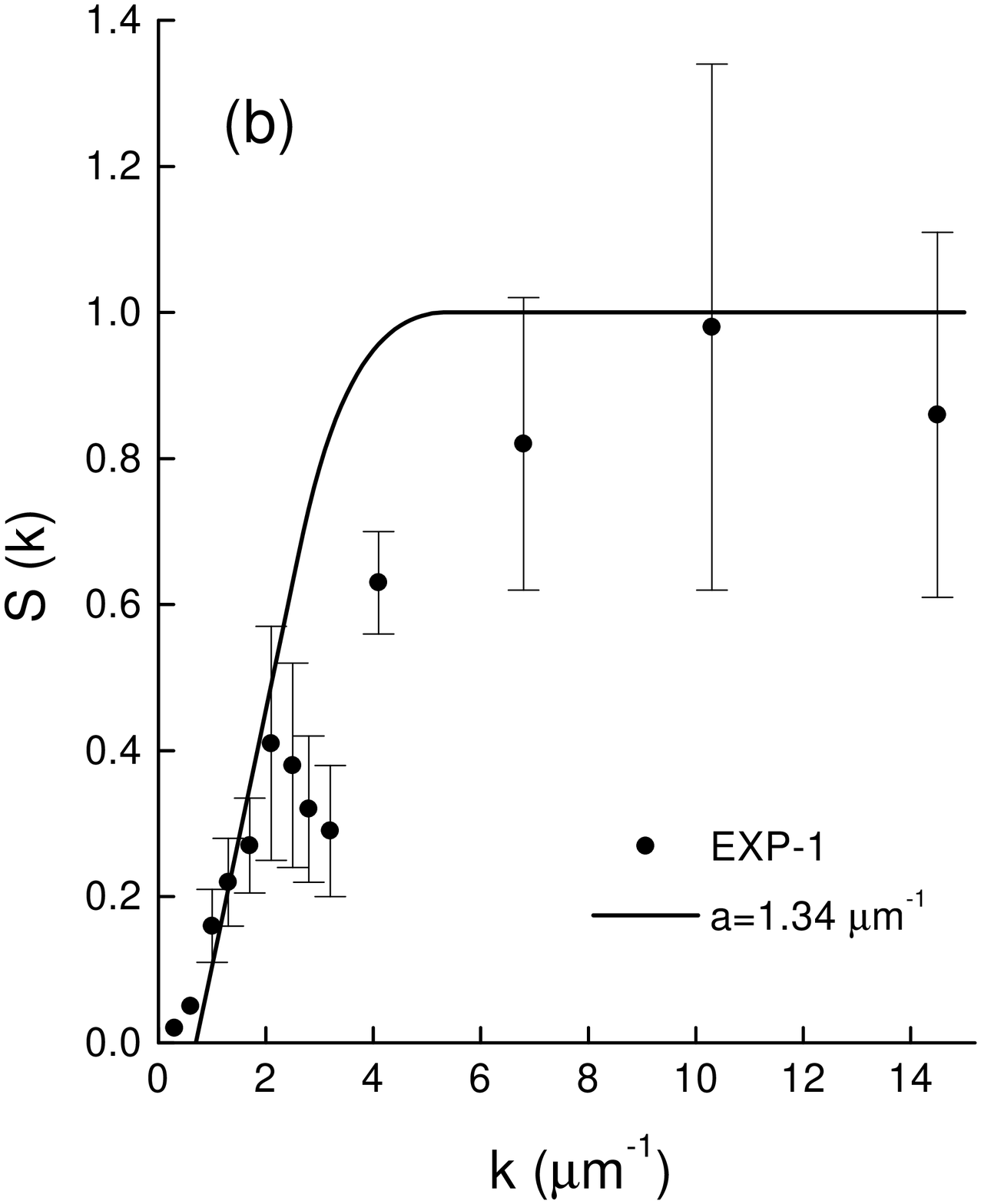,width=5.5cm} }
{\epsfig{figure=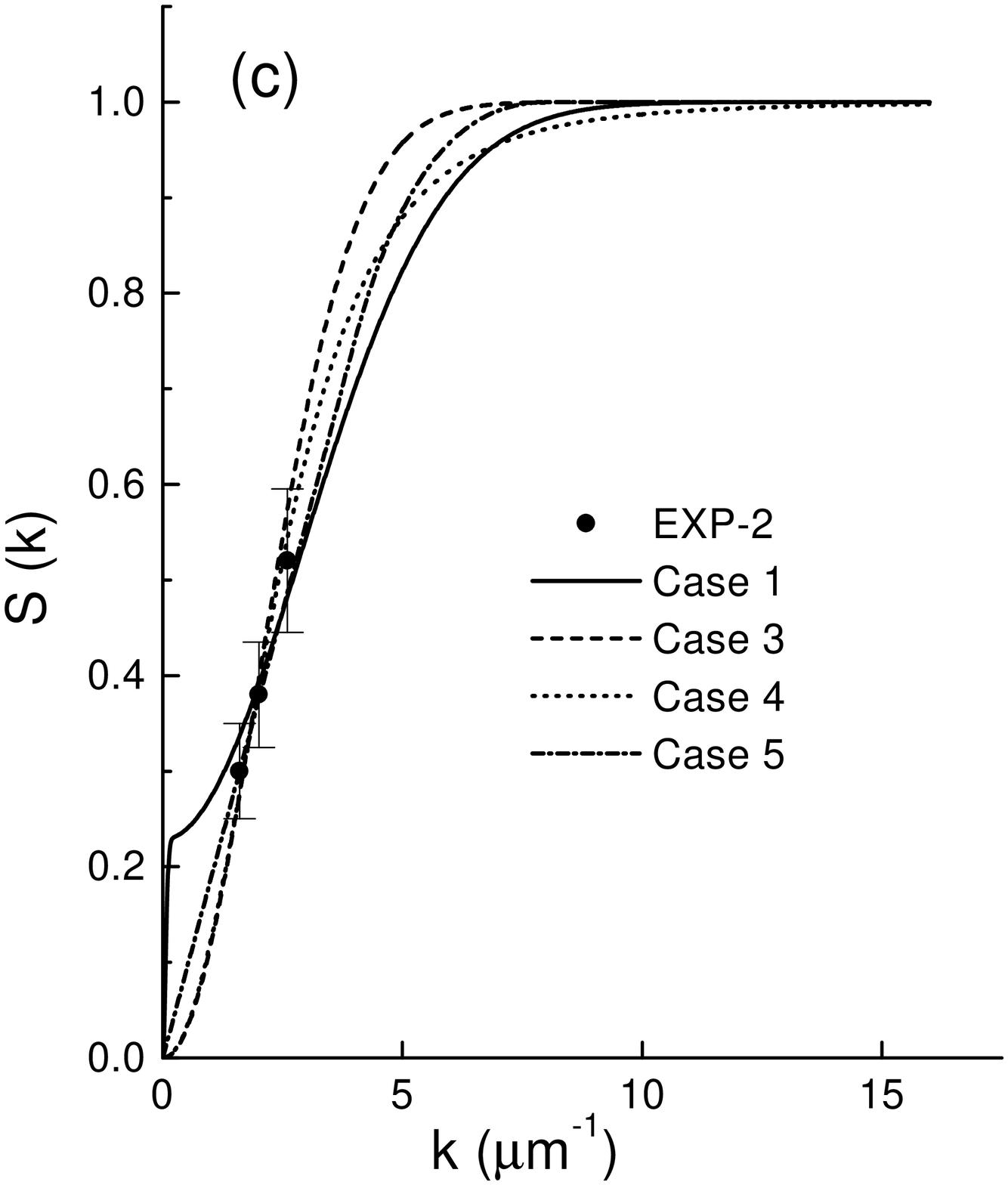,width=5.5cm} }\\
\end{tabular}
\end{center}
\caption{ The static structure factor $S(k)$ of the inhomogeneous
and homogeneous Bose gas in various cases versus the momentum $k$,
(a) in Case 1 for various values of the correlation parameter
$\beta$ as well as for the uncorrelated case (harmonic
oscillator), (b) in Case 2 for the least squares best fit value
of the parameter $a$, (c) in Cases 1,3,4 and 5 for the best fit
values of the correlation parameters. The experimental points
EXP-1 and EXP-2 are from references [2] and [1] respectively. For
the various cases see text.}
\end{figure}
%%%%%%%%%%%%%%%%%%%%%%%%%%%%%%%%%%%%%%%%%%%%%%%%%%%%%%%%%%%%%%%%%%%%%%
%%%%%% Figure 2
\begin{figure}
\begin{center}
\begin{tabular}{c}
{\epsfig{figure=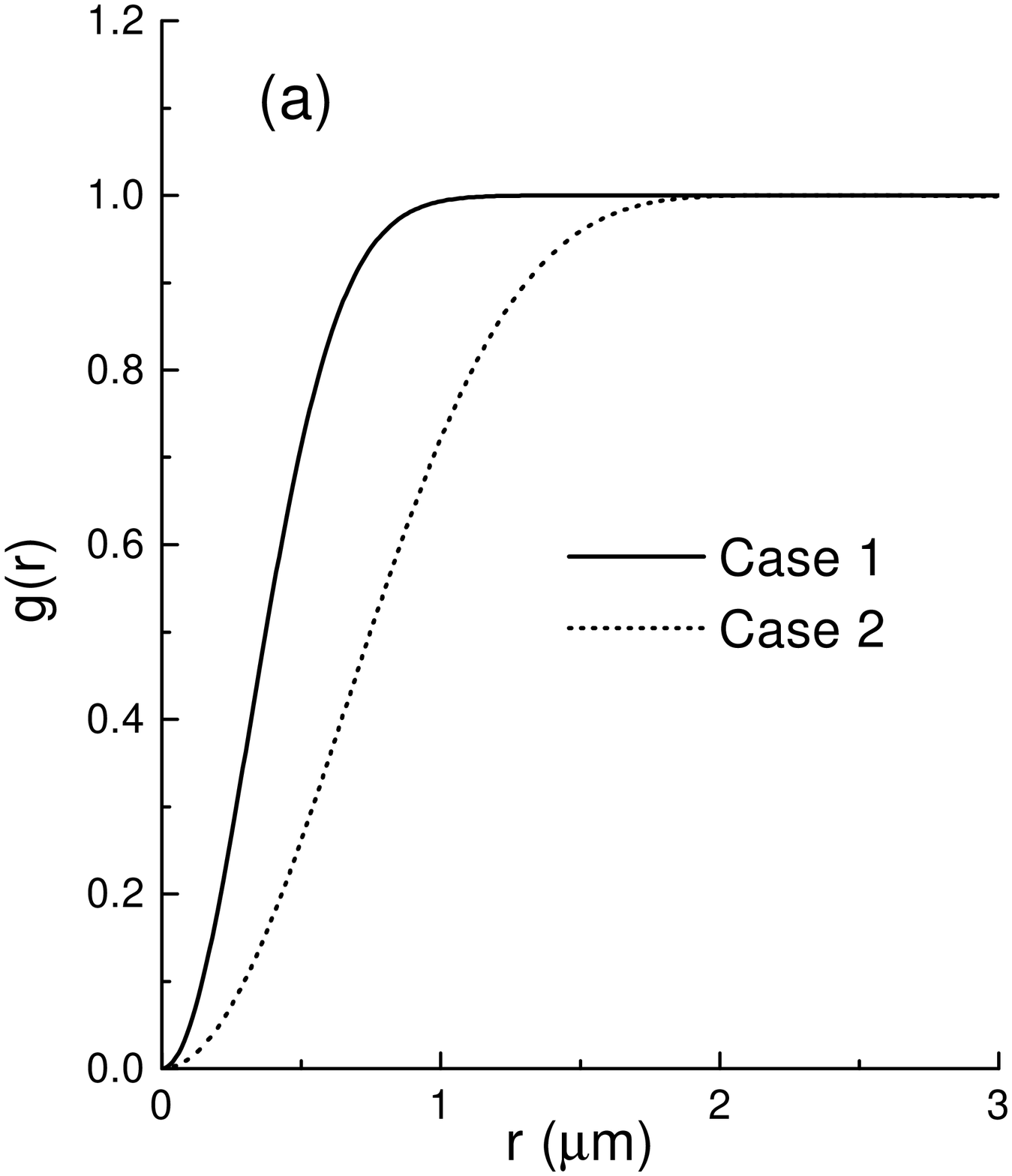,width=5.5cm} }
{\epsfig{figure=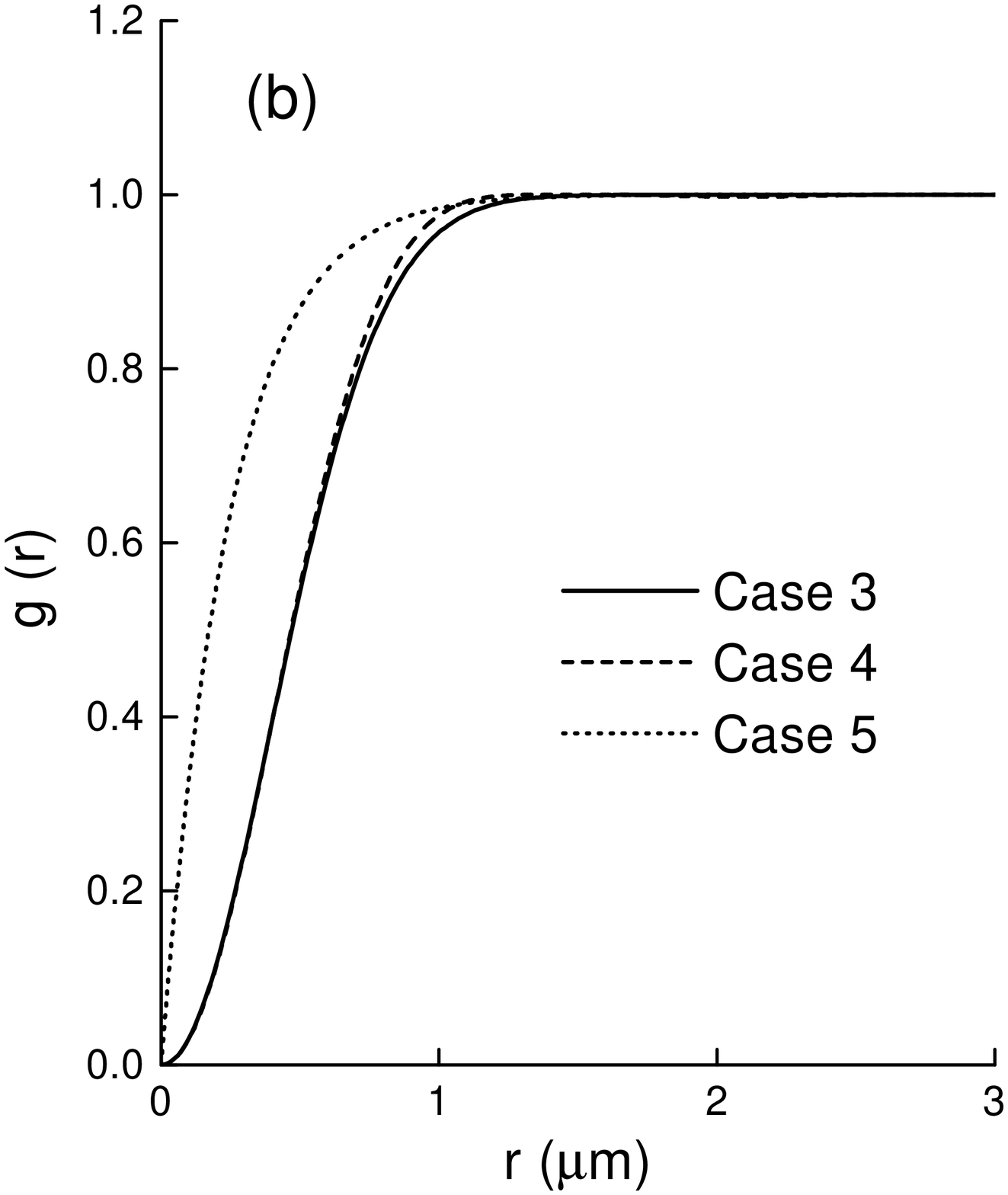,width=5.5cm} }\\
\end{tabular}
\end{center}
\caption{The radial distribution function $g(r)$ for Case 1 and
2 (a) (corresponding to inhomogeneous Bose gas), and Cases 3,4 and 5 (b)
(corresponding to homogeneous Bose gas) with the best fit values
of the correlation parameters.}
\end{figure}
%%%%%%%%%%%%%%%%%%%%%%%%%%%%%%%%%%%%%%%%%%%%%%%%%%%%%%%%%%%%%%%%%%%%%
%%%%%% Figure 3
\begin{figure}
\begin{center}
\begin{tabular}{c}
{\epsfig{figure=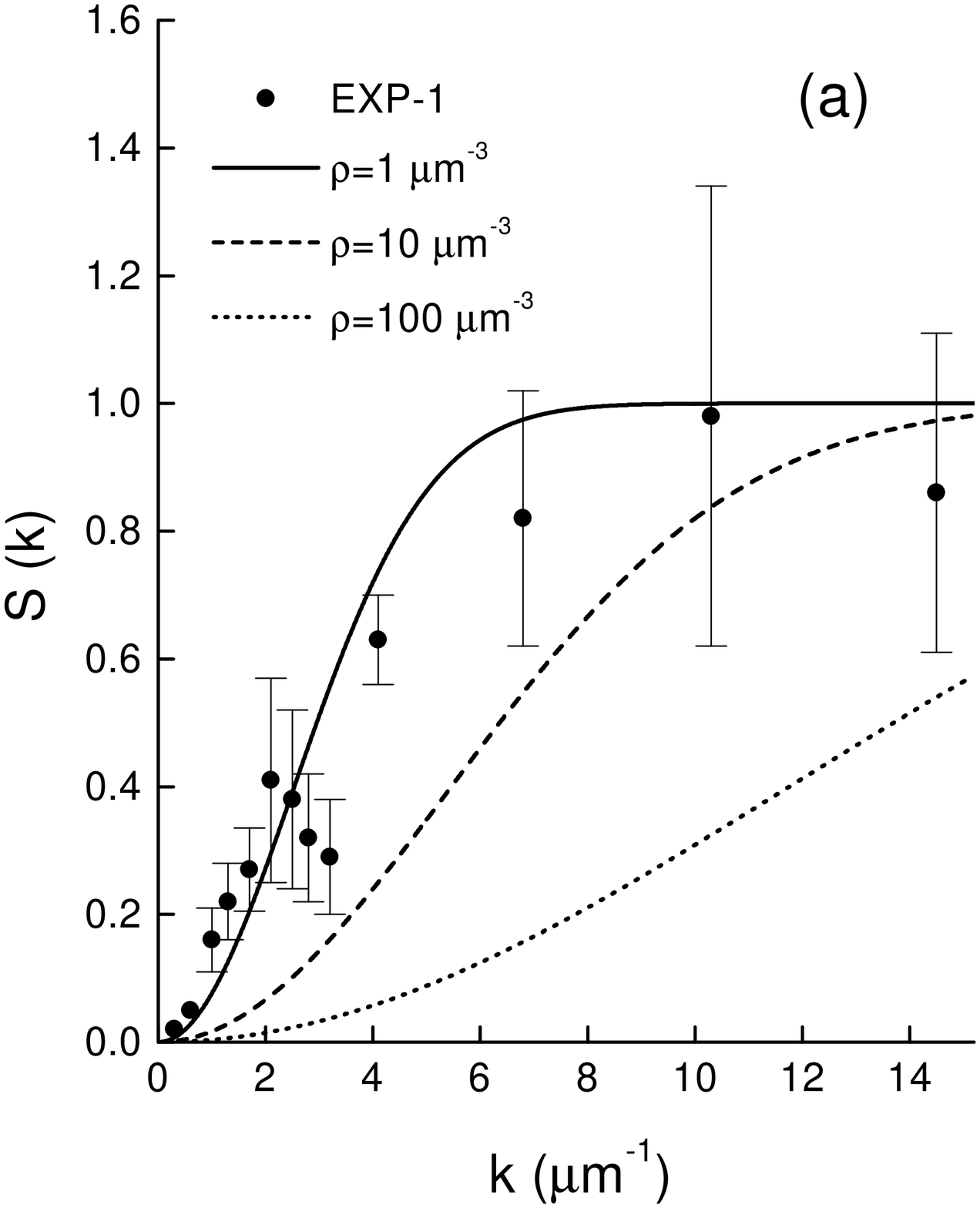,width=5.5cm} }
{\epsfig{figure=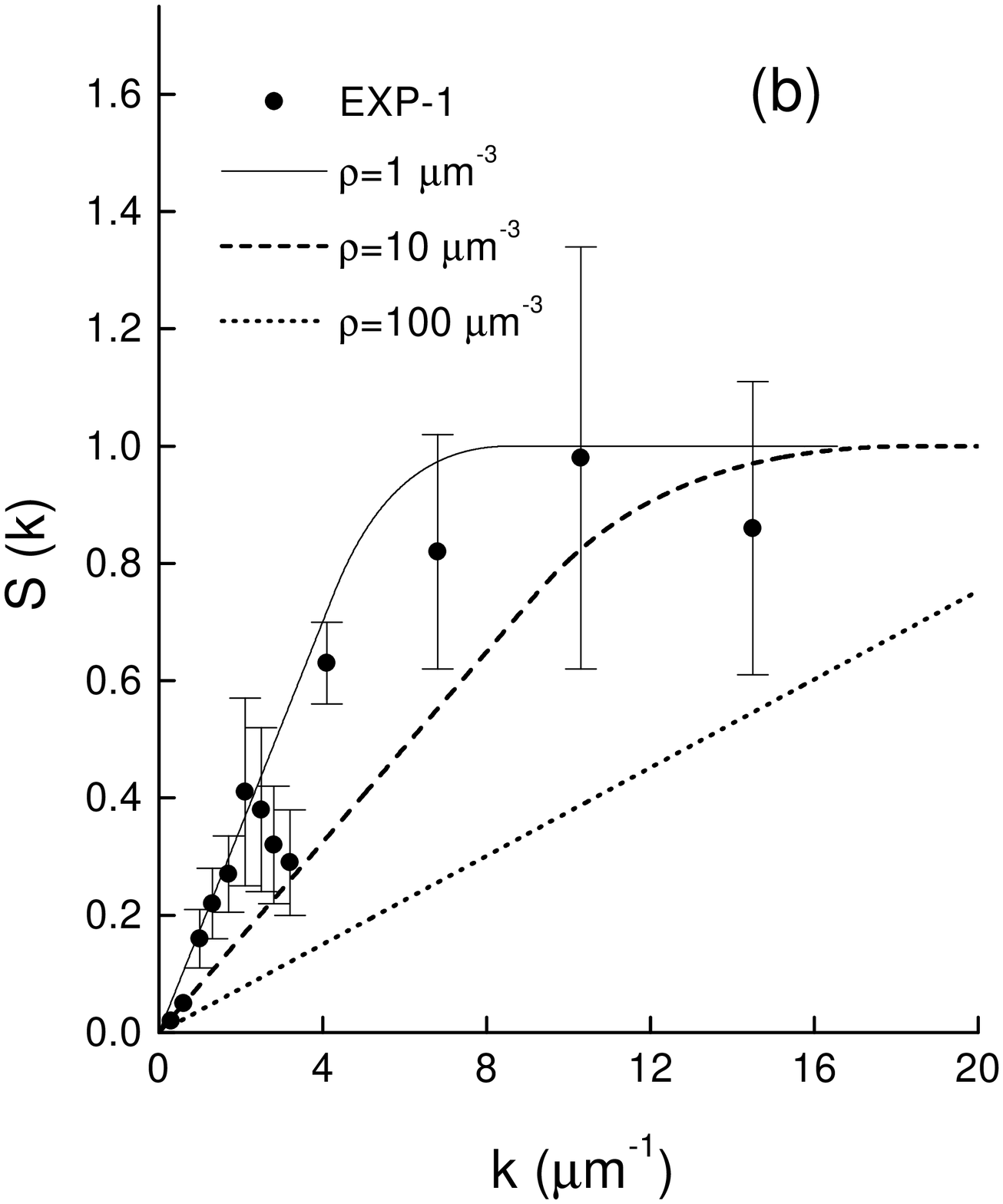,width=5.5cm} }
{\epsfig{figure=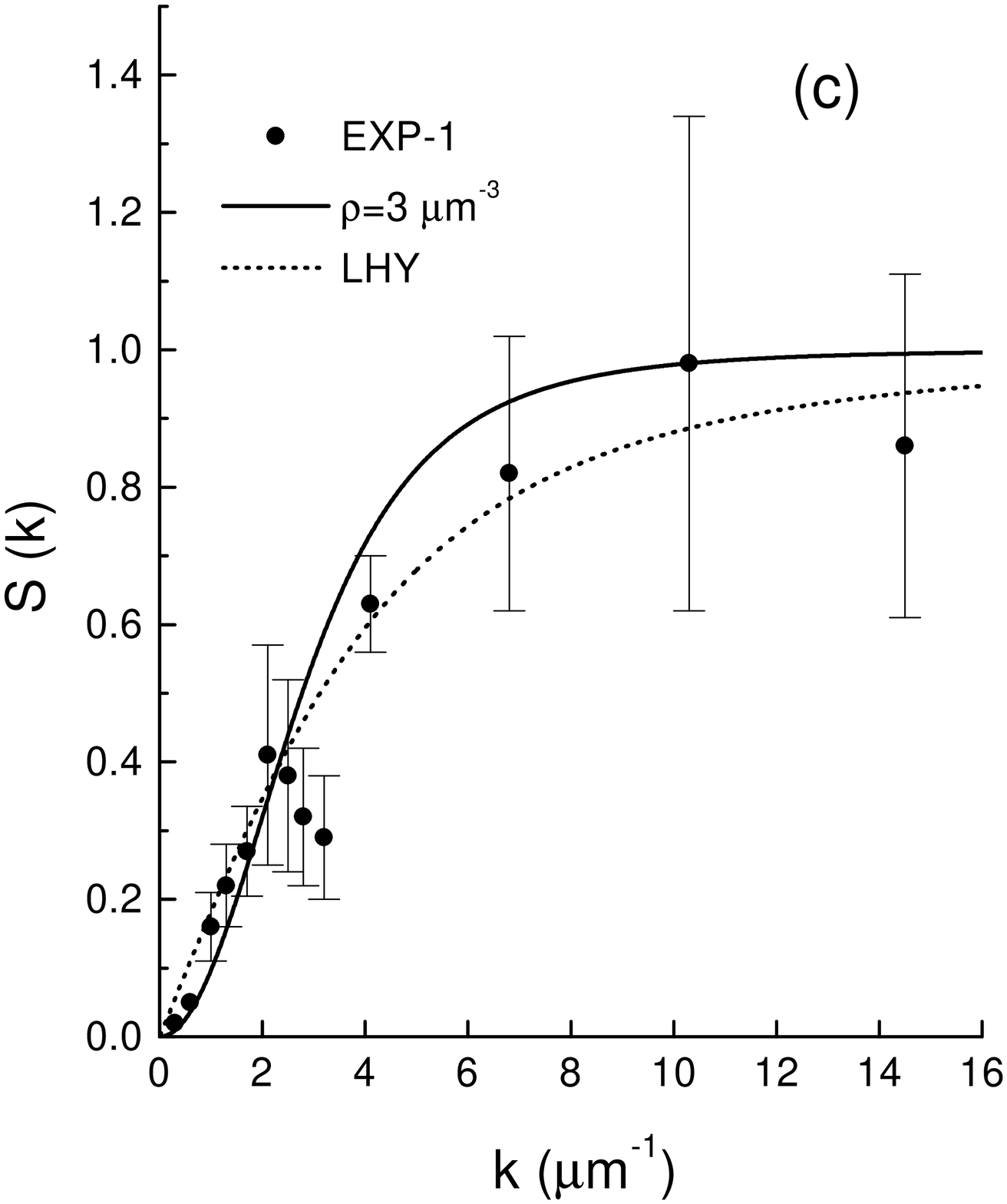,width=5.5cm} }\\
\end{tabular}
\end{center}
\caption{ The static structure factor $S(k)$ for Case 3 (a) and Case 4 (b),
for  various values of the density $\rho$.
The Case 5 (c) for the best fit
value of the density and also for the hard-sphere interaction of
Lee et.al. [11] (indicated LHY). The experimental data are from Ref. [2]}
\end{figure}

%%%%%%%%%%%%%%%%%%%%%%%%%%%%%%%%%%%%%%%%%%%%%%%%%%%%%%%%%%%%%%%%%%%
%%%%%%%%%%%%%%%%%%%%%%%%%%%%%%%%%%%%%%%%%%%%%%%%%%%%%%%%%%%%%%%%%%%%

%%%%%%%%%%%%%%%%%%%%%%%%%%%%%%%%%%%%%%%%%%%%%%%%%%%%%%%%%%%%%%%%%%%
%%%%%%%%%%%%%%%%%%%%%%%%%%%%%%%%%%%%%%%%%%%%%%%%%%%%%%%%%%%%%%%%%%%%

%%%%%%%%%%%%%%%%%%%%%%%%%%%%%%%%%%%%%%%%%%%%%%%%%%%%%%%%%%%%%%%%%%%
%%%%%%%%%%%%%%%%%%%%%%%%%%%%%%%%%%%%%%%%%%%%%%%%%%%%%%%%%%%%%%%%%%%%

\end{document}